\documentclass[twocolumn,showpacs,preprintnumbers,amsmath,amssymb]{revtex4}

\usepackage{graphicx}
\usepackage{dcolumn}
\usepackage{bm}
\usepackage{hyperref}

\begin{document}

\preprint{}

\title{ Two-component sandpile model : self-organized criticality of the second kind }

\author{Akihiro Fujihara$^{1}$}
 \email{fujihara@yokohama-cu.ac.jp}
\author{Toshiya Ohtsuki$^{2}$}
\author{Teruhiro Nakagawa$^{1}$}

\affiliation{$^{1}$Graduate School of Integrated Science, Yokohama City University, 22-2 Seto, Kanazawa-ku, Yokohama 236-0027, Japan\\
$^{2}$Field of Natural Sciences, International Graduate College of Arts and Sciences, Yokohama City University, 22-2 Seto, Kanazawa-ku, Yokohama 236-0027, Japan}

\date{\today}

\begin{abstract}
 Two-component sandpile models are investigated numerically and theoretically. Monte Calro simulations are performed to show that probability distribution functions of avalanche size and lifetime obey power laws whose exponents are approximately equal to $1.5$ and $2.0$ and the system exhibits SOC. A mean-field theory is developed to discuss the essence of the processes. We find that two-component models approach a steady critical state belonging to a different universality class from that of one-component models. Conservation of two kinds of sands at local toppling causes an infinite number of stable states which substitute for artificial boundary dissipation. Among two control parameters appearing in one-component models, therefore, a rate constant of dissipation is removed in two-component models. It is concluded that the more conserved quantities result in the less control parameters and a novel class of SOC.
\end{abstract}

\maketitle

 According to a vast amount of studies on fractals, $1/f$ noise, and so on, it has been reported that a lot of nonequiliblium systems in nature exhibit power laws. However it has not been deeply understood why a huge variety of distributions spontaneously go to power laws. Self-organized criticality(SOC) proposed by Bak \textit{et al.}\cite{BTW1987, BTW1988, Jensen, Sornette} has provided one reasonable understanding of the emergence of power laws through sandpile models. The models are cellular automata evolving in threshold dynamics. The domino effect induces a sequence of topplings called an avalanche, whose distributions of magnitude and lifetime obey power laws without tuning any parameters. A number of works on SOC have been done numerically. Several analytical results have also been obtained. Dhar\cite{Dhar1990, Dhar1999} has found some exact solutions in abelian sandpile models and has shed light on the statics of the models, such as the number of total recurrent states and hight correlation functions. However, the dynamics of sandpile models is still obscure, that is, the power-law distributions of avalanches still have been unable to be derived analytically. 

 Though sandpile models seem to have no control parameters by appearances, Vespegnani and Zapperi\cite{VZ1997, VZ1998} have found that the models do have hidden control parameters: a slow dissipation rate and a slower addition rate of sands. To explain this, they have introduced a mean-field theory of rate equations,
\begin{eqnarray}
\frac{\partial}{\partial t}\rho_{\kappa} = f_{\kappa}(\rho_{a}, \rho_{c}, \rho_{s}),\hspace{0.5cm} \kappa = a, c, s
\end{eqnarray}
where $\rho_{a}$, $\rho_{c}$, and $\rho_{s}$ are the probability densities of active, critical, and stable states, respectively. Imposing the conservation law of the number of sands in a local toppling rule, it is concluded that the models become critical in the double limit 
\begin{eqnarray}
\epsilon, \delta \to 0, \hspace{5mm} \rho_{a}=\delta/\epsilon \to 0,\label{eq:double_limit}
\end{eqnarray}
where $\epsilon$ is a dissipation rate and $\delta$ an addition rate. In other words, $\rho_{a}$ is an order parameter and $\epsilon$, $\delta$ are control parameters. SOC is achieved by rough tuning of the "unapparent" parameters $\epsilon$, $\delta$ around zero. Most of SOC models\cite{Manna1991, DS1992} belong to this "universality" class. We call this class \textit{SOC of the first kind}. It seems interesting whether other kind of SOC exists, or equivalently, whether the condition of the double limit (\ref{eq:double_limit}) are strictly the necessary condition of SOC? In this letter, we give an answer to these questions.

 As mentioned above, the conservation law of the local toppling rule plays a key role for SOC. Tsuchiya and Katori\cite{TK2000} have proved rigorously that SOC breaks down in abelian sandpile models with nonconservative toppling of sands. Here, we pay attention to the number of conservation laws. Does anything new happen if the number of conservation laws is increased? How about the robustness of SOC? To answer these questions, we consider \textit{two-component sandpile models} which deal with two kinds of sands. This means the models have two kinds of conservation laws. Firstly, numerical simulations are carried out to check power-law behavior. Secondly, a mean-field theory is developed to examine the essence of the processes.

\begin{figure}
\begin{center}
\resizebox{80mm}{!}{\includegraphics{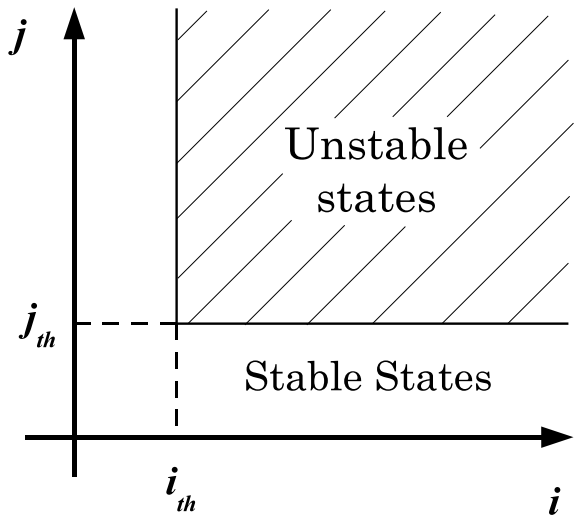}}
\caption{Stable and unstable states in the rule (b).}
\label{fig:sustates}
\end{center}
\end{figure}

 A two-component sandpile model is a cellular automaton defined on a regular lattice. A pair of two non-negative integers $\boldsymbol{h}(\boldsymbol{x}) \equiv (i, j)$ is assigned to each site $\boldsymbol{x}$ on the lattice, where $i,j (\ge 0)$ denote the numbers of sand $A$ and $B$. At each time step, one unit of sand $A$ or $B$ is added to a randomly chosen site at a relative ratio $(0 <) r_{AB} (< 1)$ of sands $A$ to sands $B$. Several types of toppling rule can be adopted. For example, a toppling occurs when (a) $i \ge i_{th}$ \textit{or} $j \ge j_{th}$, (b) $i \ge i_{th}$ \textit{and} $j \ge j_{th}$, or (c) $i+j \ge k_{th}$ $(i, j > 0)$ where $i_{th}$, $j_{th}$, and $k_{th}$ are certain threshold values. If the number of sands at one site reaches threshold, sands $A$ and $B$ on the site topple one by one to randomly chosen nearest neighbor sites until the number of sands at the site becomes less than threshold. In these rules, a sand $A$ and a sand $B$ topple jointly. For example, if a site has $n$ grains of sands $A$ and no sands $B$, and then $m$ ($\le n$) new grains of sands $B$ are added, it ends up with $n-m$ grains of sands A and no sands $B$. These rules satisfy two kinds of conservation laws of local toppling. A series of topplings (an avalanche) continue unless the numbers of sands at all sites become less than threshold. Among three rules of toppling, the \textit{and} rule (b) is the most interesting because of the presence of an \textit{infinite} number of stable states. In the rule (b), stable states $\boldsymbol{h}_{s}$ and unstable states $\boldsymbol{h}_{u}$ are defined by 
\begin{eqnarray}
\boldsymbol{h}_{s} &=& \{ (i,j) \; | \; 0 \le i < i_{th} \; \textrm{or} \; 0 \le j < j_{th} \}, \\
\boldsymbol{h}_{u} &=& \{ (i,j) \; | \; i \ge i_{th} \; \textrm{and} \; j \ge j_{th} \}.
\end{eqnarray}
and there are infinite stable and unstable states as illustrated in Fig.~\ref{fig:sustates}.
These stable states can absorb all the added sands. This means avalanches can stop without introducing artificial boundary dissipation, which is indispensable in one-component models. Rules (a) and (c) are not capable of removing boundary dissipation because the number of stable states is finite under these rules. Hereafter, we consider only the rule (b) and treat threshold $i_{th} = j_{th} = 1$ and periodic boundary conditions. 

\begin{figure*}
\begin{tabular}{cc}
\resizebox{80mm}{!}{\includegraphics{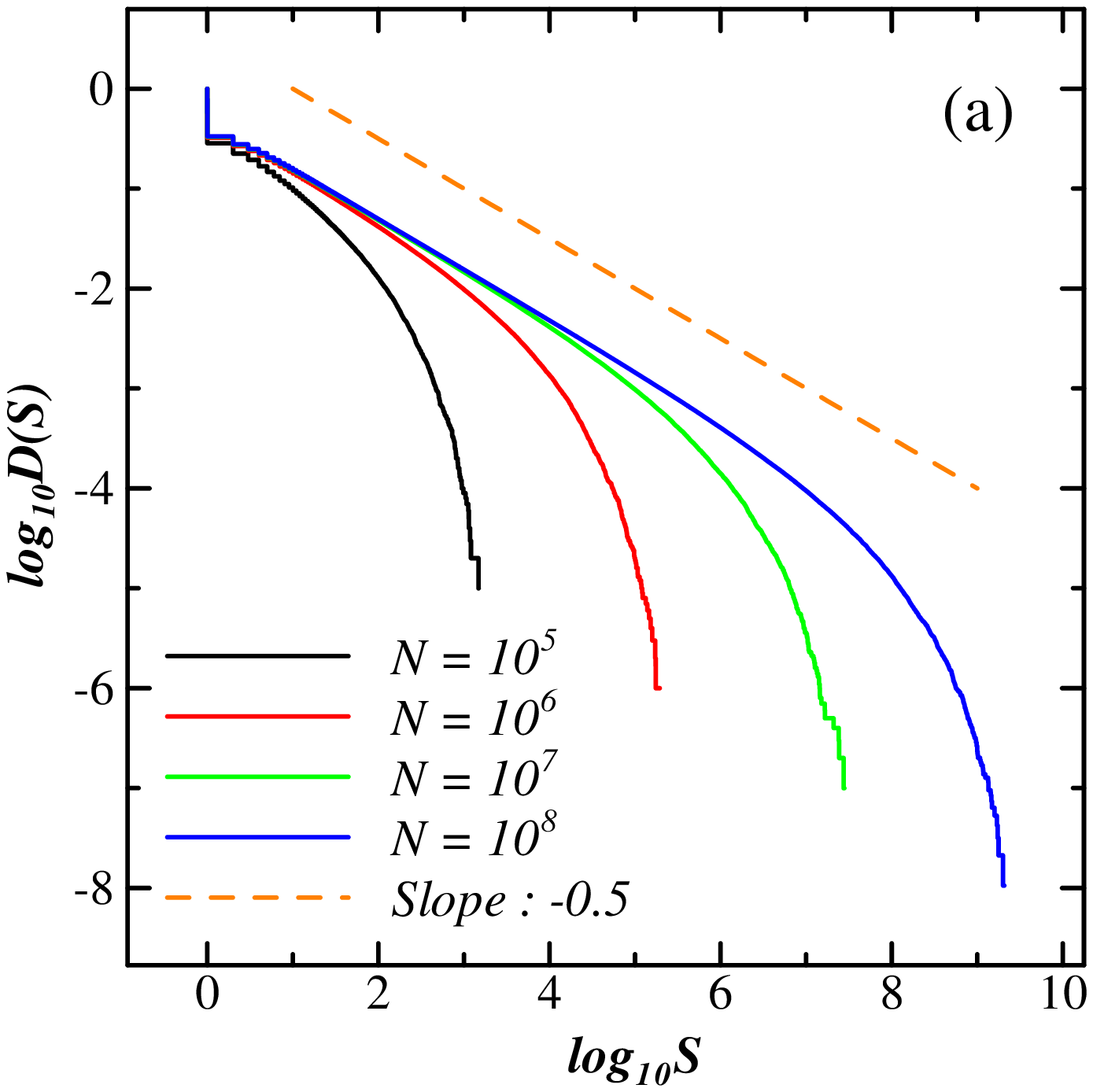}}
\resizebox{80mm}{!}{\includegraphics{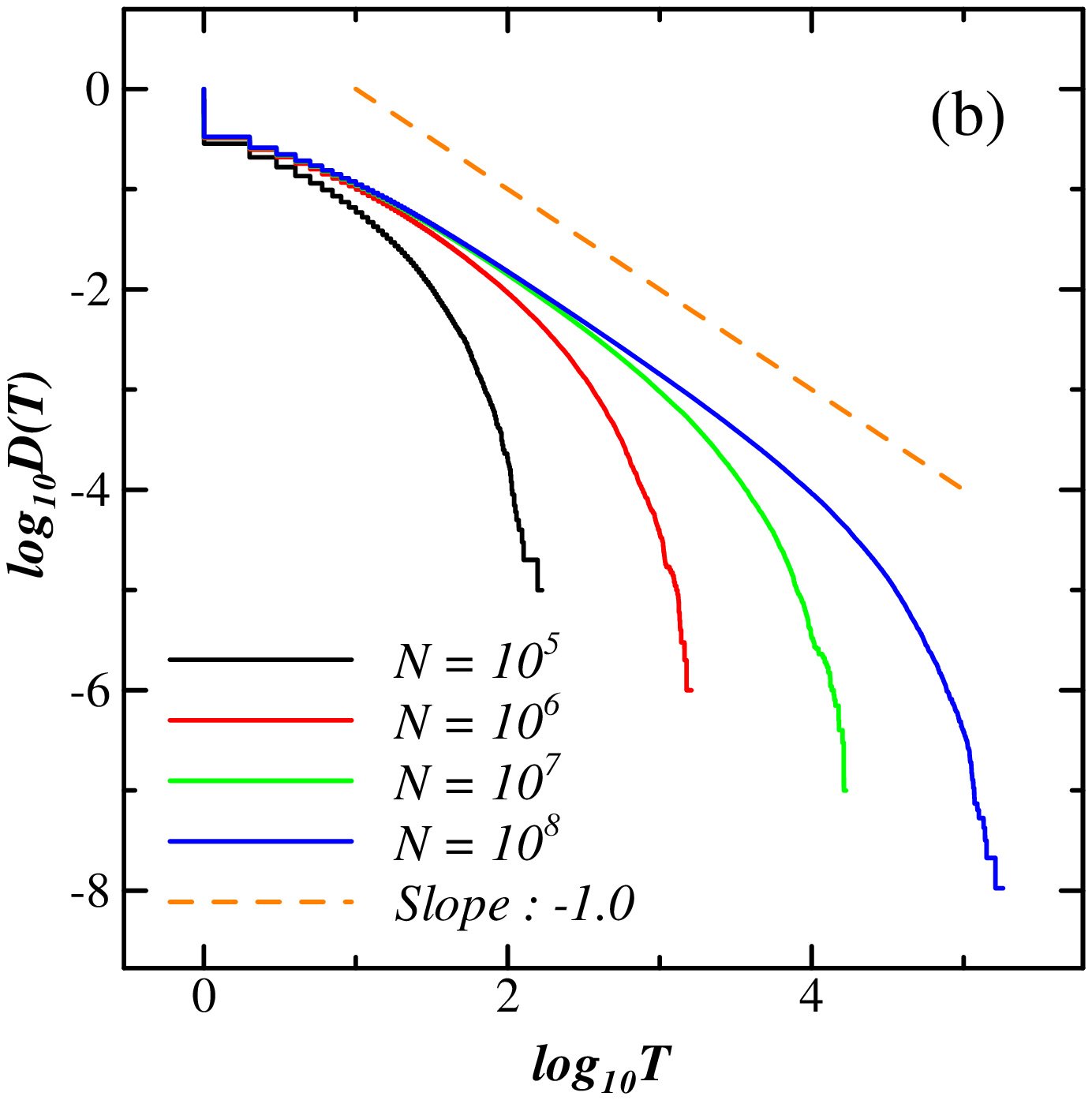}}
\end{tabular}
\caption{Double logarithmic plots of CDFs of avalanche size (a) and lifetime (b). The lattice size $L=10^{4}$ and the number of added sands $N=10^{5}$, $10^{6}$, $10^{7}$, and $10^{8}$, respectively. Sands $A$ and $B$ are added at equal rate $r_{AB} = 0.5$. }
\label{fig:2dist}
\end{figure*}

 Distribution functions of avalanche size $S$ and lifetime $T$ are calculated numerically, where $S$ is defined by the total number of topplings in one avalanche and $T$ by the total time steps of simultaneous updates of topplings. Cumulative distribution functions (CDFs) of avalanche size $D(S)$ and lifetime $D(T)$ in $1-$dimensional lattices are shown in Figs.~\ref{fig:2dist}.
 We find that the CDFs obey power laws $D(S) \sim S^{-\tau}$ and $D(T) \sim T^{-\alpha}$ where the exponents are approximately $\tau \simeq 0.5$ and $\alpha \simeq 1.0$. Equivalently, probability distribution functions (PDFs) are represented as $P(S) \sim S^{-(1+\tau)}$ and $P(T) \sim S^{-(1+\alpha)}$. These exponents $1+\tau \simeq 1.5$ and $1+\alpha \simeq 2.0$ are very close to the critical exponents of branching processes\cite{Harris}. We carried out simulations in $2-$ and $3-$dimension and obtained almost the same CDFs as those in $1-$dimension. These results imply the model has meanfield-like characteristics\cite{Alstrom1988}. It is also found that varying $r_{AB}$ from $0.5$ to $0.1$ hardly affects power-law tails. This means the model ends up with going to SOC states robustly in the long time limit without tuning $r_{AB}$. It becomes evident that the two-component sandpile models exhibits SOC. It should be noted that because of the absence of boundary dissipation, the power-law tails become extended infinitely as the number of added sands $N$ increases even in a finite system. 

 Next, we construct a mean-field theory of two-component sandpile models and compare with that of one-component models\cite{VZ1997, VZ1998}. We introduce the sets of probability densities of stable states $\boldsymbol{X}$ and unstable states $\boldsymbol{Z}$ as
\begin{eqnarray}
\boldsymbol{X} &=& \{ X_{(i,j)} \; | \; 0 \le i < 1 \; \textrm{or} \; 0 \le j < 1 \}, \\
\boldsymbol{Z} &=& \{ Z_{(i,j)} \; | \; i \ge 1 \; \textrm{and} \; j \ge 1 \}.
\end{eqnarray}
Rate equations for $\boldsymbol{X}$ and $\boldsymbol{Z}$ are given by an infinite system of first-order nonlinear ordinary differential equations. However, it is difficult to handle the infinite degrees of freedom. In order to avoid the difficulty, we define the following reduced six variables
\begin{eqnarray}
\boldsymbol{X^{*}} = \{X_{0}, X_{A}, X_{B}\}, \hspace{5mm} \boldsymbol{Z^{*}} = \{Z_{0}, Z_{A}, Z_{B}\}
\end{eqnarray}
with
\begin{eqnarray*}
X_{0} \equiv X_{(0,0)}, \hspace{12mm} Z_{0} \equiv \sum_{i=1}^{\infty} Z_{(i,i)}, \\
X_{A} \equiv \sum_{i=1}^{\infty} X_{(i,0)}, \hspace{5mm} Z_{A} \equiv \sum_{i=2}^{\infty} \sum_{j=1}^{\infty} Z_{(i,j)}, \\
X_{B} \equiv \sum_{j=1}^{\infty} X_{(0,j)}, \hspace{5mm} Z_{B} \equiv \sum_{i=1}^{\infty} \sum_{j=2}^{\infty} Z_{(i,j)}.
\end{eqnarray*}
For obtaining a closed set of equations for $\boldsymbol{X^{*}}$ and $\boldsymbol{Z^{*}}$, we assume Poisson distributions for $X_{(i,0)}$ and $X_{(0,j)}$ as
\begin{eqnarray}
\frac{X_{(i,0)}}{X_{A}} &=& \frac{\mu_{A}^{i-1}}{(i-1)!}e^{-\mu_{A}} \hspace{5mm} (i \ge 1), \\
\frac{X_{(0,j)}}{X_{B}} &=& \frac{\mu_{B}^{j-1}}{(j-1)!}e^{-\mu_{B}} \hspace{5mm} (j \ge 1)
\end{eqnarray}
where $\mu_{A} \equiv \delta_{A}t$, $\mu_{B} \equiv \delta_{B}t$ are mean values of each Poisson distribution at time $t$ and $\delta_{A}$, $\delta_{B}$ rate constants of additions of each sand. Rate equations for the reduced variables read
\begin{eqnarray}
\frac{d X_{0}}{dt} &=& -2 X_{0}(Z_{0}+Z_{A}+Z_{B}) + Z_{0} \nonumber \\
		   & & - (\delta_{A}+\delta_{B})X_{0}, \label{eq:X0} \\
\frac{d X_{A}}{dt} &=& (X_{0}-X_{A})(Z_{0}+Z_{A}+Z_{B}) + Z_{A} \nonumber \\
		   & & + \delta_{A}X_{0} - \delta_{B}X_{A}, \\
\frac{d X_{B}}{dt} &=& (X_{0}-X_{B})(Z_{0}+Z_{A}+Z_{B}) + Z_{B} \nonumber \\
		   & & + \delta_{B}X_{0} - \delta_{A}X_{B}, \\
\frac{d Z_{0}}{dt} &=& (\alpha_{A}X_{A}+\alpha_{B}X_{B}-2Z_{0})(Z_{0}+Z_{A}+Z_{B}) - Z_{0} \nonumber \\
		   & & + \alpha_{B}\delta_{A}X_{B} + \alpha_{A}\delta_{B}X_{A}, \\
\frac{d Z_{A}}{dt} &=& \{ (1-\alpha_{A})X_{A}+Z_{0} \}(Z_{0}+Z_{A}+Z_{B}) - Z_{A} \nonumber \\
		   & & + (1-\alpha_{A})\delta_{B}X_{A}, \\
\frac{d Z_{B}}{dt} &=& \{ (1-\alpha_{B})X_{B}+Z_{0} \}(Z_{0}+Z_{A}+Z_{B}) - Z_{B} \nonumber \\
		   & & + (1-\alpha_{B})\delta_{A}X_{B} \label{eq:ZB}
\end{eqnarray}
where $\alpha_{A}(t) \equiv X_{(1,0)}/X_{A} = \exp(-\delta_{A} t)$ and $\alpha_{B}(t) \equiv X_{(0,1)}/X_{B} = \exp(-\delta_{B} t)$.

\begin{figure}
\begin{center}
\resizebox{80mm}{!}{\includegraphics{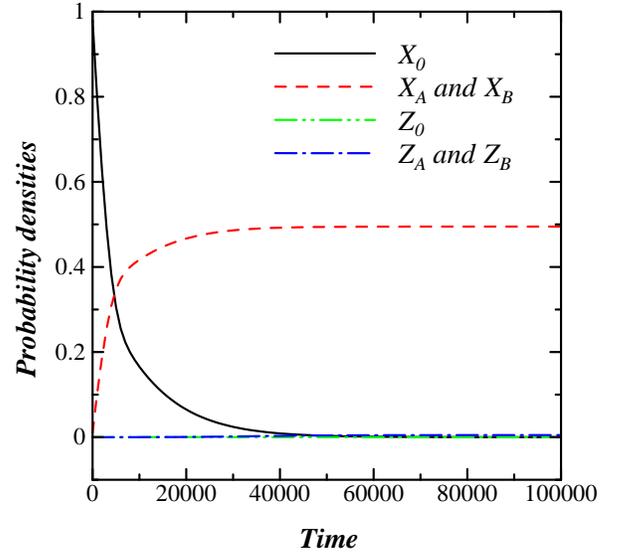}}
\caption{Time evolution of the reduced six variables at $\delta_{A}=\delta_{B}=0.0001$. Initial values are set to $X_{0}=1.0$ and $X_{A}=X_{B}=Z_{0}=Z_{A}=Z_{B}=0.0$. At the steady state, $X_{0}=Z_{0} \simeq 0$, $X_{A}=X_{B} \simeq 0.5$, and $Z_{A}=Z_{B} \simeq \sqrt{\delta_{A}}/2 = 0.005$.}
\label{fig:meanfield}
\end{center}
\end{figure}

 Numerical solutions of the equation system (\ref{eq:X0})-(\ref{eq:ZB}) are plotted in Fig.~\ref{fig:meanfield}. 
It is found that in the long time limit, the system goes to a steady state which is supposed to be SOC. The steady state is able to be examined analytically. When $t \to \infty$, then $\alpha_{A}, \alpha_{B} \to 0$ and $X_{0}, Z_{0} \to 0$. Therefore the following relations are derived from  Eqs.~(\ref{eq:X0})-(\ref{eq:ZB}).
\begin{eqnarray}
X_{A} = \frac{Z_{A}}{Z_{A}+Z_{B}+\delta_{B}}, \hspace{5mm} X_{B} = \frac{Z_{B}}{Z_{A}+Z_{B}+\delta_{A}}.
\end{eqnarray}
Suppose that $Z_{A}+Z_{B} \gg \delta_{A}, \delta_{B}$ and $Z_{A} : Z_{B} = \delta_{A} : \delta_{B}$, we obtain
\begin{eqnarray}
Z_{A} = \frac{\sqrt{2}\delta_{B}^{\frac{1}{2}}\delta_{A}^{\frac{3}{2}}}{(\delta_{A}+\delta_{B})^{\frac{3}{2}}}, \hspace{5mm} Z_{B} = \frac{\sqrt{2}\delta_{A}^{\frac{1}{2}}\delta_{B}^{\frac{3}{2}}}{(\delta_{A}+\delta_{B})^{\frac{3}{2}}}. \label{eq:critical}
\end{eqnarray}
Equations (\ref{eq:critical}) show that the model goes to a critical state $(Z_{A}=Z_{B}=0)$ in the limit $\delta_{A}, \delta_{B} \sim \delta \to 0$. Out of two conditions (\ref{eq:double_limit}) for SOC of the first kind, our model can successfully remove one of them with respect to the control parameter of dissipation $\epsilon$. Furthermore, the order parameters behave as $Z_{A}, Z_{B} \sim \delta^{1/2}$ when the control parameters go to zero $\delta \to 0$. These two characteristic features of two-component sandpile models, that is the existence of the steady state without dissipation and the asymptotic behavior to the critical state in $\delta \to 0$, distinctively differ from usual SOC models and are strong evidences that the model belongs to different universality class. Consequently, we are successfully able to find a new mechanism and class of SOC. So, we call this \textit{SOC of the second kind}. 

 We have investigated two-component sandpile models with the rule (b) and periodic boundary conditions mainly on $1-$dimensional lattices. Extention to two components is essential to generate an infinite number of stable states which substitute for boundary dissipation. Therefore, the model becomes more natural because the dissipation is usually introduced artificially only for stopping avahanches. Simulation results indicate the model shows SOC behavior. The mean-field theory confirms that the model goes to the critical steady state belonging to a different universality class from one-component models. Here, only rough-tuning of rate constants of additions $\delta$ is enough to induce SOC. We can successfully construct a new SOC model without introducing any artificial dissipation mechanisms. Consequently, the two conditions (\ref{eq:double_limit}) are not the necessary condition for SOC, which is the answer to the question mentioned previously. 

 It is interesting to summarize SOC models from the standpoint of the number of conserved quantities. Models that have no conserved quantities, such as contact processes, show power laws strictly on the critical point. Therefore, fine-tuning of control parameters is required. Models that have one conserved quantity, such as one-component sandpile models, become SOC in rough-tuning of \textit{two} control parameters, an addition and a dissipation rates of sands, around zero. Two-component sandpile models with two conserved quantities show SOC in rough-tuning of only \textit{one} control parameter, an addition rate of sands, around zero. It could be concluded that the more conservation laws result in the less control parameters. 

 Sandpile models are considered to have some relevance to power laws of earthequake magnitude distributions, called Guthenberg-Richter law\cite{Turcotte1999}. One-component sandpile models have to introduce some artificial boundary dissipations in order to stop avalanches. However, the surface of the earth has no apparent boundary. At this point, it is more natural to apply two-component sandpile models to earthequake magnitude distributions. Furthermore, an earthquake could be triggered by multiple physical quantities, such as elastic energy and stress of earth's crust. If an earthquake takes place only when both the energy and stress reach threshold simultaneously, the distributions of earthquakes would be well described by the SOC of the second kind. In this letter, we express the process in terms of a sandpile. Obviously, sands are merely symbols and could be any kind of substances which trigger the dynamics. In addition to earthquakes, therefore, a wide variety of SOC phenomena triggered by multiple factors would belong to SOC of the second kind. Future studies will make clear these points.

\end{document}